# METHODS OF EXECUTABLE CODE PROTECTION

*ANTON PETROV, Ph.D., Associate Professor*
*Volodymyr Dahl East Ukrainian National University, Luhansk, Ukraine*

**Summary**
The article deals with the problems in constructing a protection system of executable code. The techniques of breaking the integrity of executable code and ways to eliminate them are described. The adoption of virtual machine technology in the context of executable code protection from analysis is considered. The substantiation of the application of virtual machines as the best way to oppose the analysis of executable code is made. The protection of executable code by transferring the protected code in a virtual execution environment is considered. An efficient implementation of the method is proposed.

**Keywords:** program code, executable code protection, virtual machine, interpretation, disassembling.

## Introduction

The information processing in computer systems and networks is the problem of to-day (and it'll be still urgent for a long time). In particular, the task of code and processed data protection is still not solved. The main objectives of the code protection program are:
1. Preventing unauthorized changes in the program behavior algorithm.
2. Protection of processed data against an illegitimate user.
3. Ensuring of the software protection from unauthorized copying.

Ideal protection in the field of software is still not created, although there is a war for decades between protection creators and hackers. Invariably the last are the victors, yielding the palm to protection creators only by chance. Under the protection the authors mean a unit of executable code, which counteracts (successfully or not) the illegal use of the program. This can be a protection from copying of storage device, input of the serial registration number of the program, means that limits the maximum number of licenses in the network and so on.

Next, let us consider the methods and procedures by which the above objectives are achieved. Note that none of these goals at the moment is reached and the corresponding scientific and technological problems are not fully resolved. Since it makes no sense to reinvent the protection against unknown threats, not knowing where, how, and when the next attack will be made. We shall consider the protection means, starting with the attack tools. In other words , we will make the analysis starting with the threat and moving to the protection means, speaking first on the side of the attacker, and then on the side of information security specialists.

**1. The urgency of the program code protection**

There is a huge field of activity for hackers in the field of program behavior change algorithm. The fact that the majority of program code is in the open form, i.e. have a view of either the machine code or some intermediate bytecode (assembler program) executed by the virtual machine.

The main problem faced by programmers creating protection code (that portion of the code that makes the decision to open or close access to the main code, in other words, performs the functions of identification, authentication and authorization) is a problem of potential capacity of analysis and correction of definite binary code.  Moreover, this vulnerability, according to the authors, is one of the most problematic in the field of software protection.

Today, more than one solution of application code encryption, allowing to secure it from disassembling, and connect to the dongle, for example, Guardant is widely known

Take the most popular method of encryption. Encode important sections of code so that they can be correctly decoded only through legally obtained key. Keyless application ceases to be operational, or work during the evaluation version. Encoding algorithm will be a test of legitimate use. Conditional jump IF "decoded_correct " THEN ... - It's just a formality that will eliminate an access violation when

Performing an incorrectly decoded area. Application crypto remains vulnerable to attack when an intruder is buying a licensed copy and running it removes from memory dump decoded version. Consequently, immediately after the execution of the decoded area application requires encode it again, or move back to pre- stored encrypted portion. Cracker will have the only opportunity to catch the moment when in  the memory the decoded portion is executed, keep working application code and  go on acting in such a way, eventually collecting application into a whole piece by piece , but this is a manual hard work , especially if you provide a large number of small areas of code, with its consistent decoding. The fight against this kind of attack is the largest possible number of coded sections, replacing them in each new version of the system [1].

There are created enough of disassemblers for machine code[2], able to provide a program in a convenient form (in assembly language) for review. Thus, by changing at least one machine instruction , we obtain the following possible outcomes :
- inefficient code;
- postponed inoperative code ( from a few seconds to several hours or days);
- misbehaving code;
- deliberately misbehaving code;

The consequences of such events are obvious. And the most pitiable result we expect in the latter case. Depending on the domain software, it may be the loss of large sums of money in the calculations, allowing system access to illegitimate user, and so on - the list is endless.

**2. Ensuring the integrity of the code and data security**

There are several security solutions for the PC.

First, use the executable packers such as UPX [3]. However, it stops only an inexperienced cracker. Unpacker signature attached at the beginning of the file, is easily recognized both manually and automatically, hence, reverse of the process of bringing the code into the original loose form is easily performed.

Secondly, special tools such as, ASProtect - the system of software protection of applications from unauthorized copying, designed for quick implementation of application protection functions, especially targeted for software developers are often used [4]. Such protection is often written by the developer manually, and as a consequence, its level is even worse, because protection writing requires at least minimal qualifications in the field of software protection writing. The major schemes of such tools can be seen in Fig. 1.

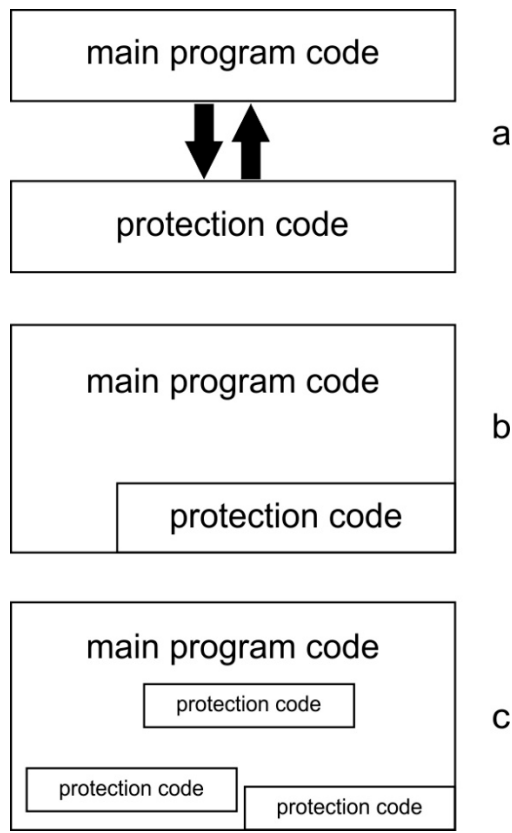

**Figure. 1. Schemes of executable code protections tools. (A - protection outside of the program , B - protection within the program, C - protection dispersed within the program)**

The main drawback of such fatal vulnerability protection is the point of basic code contact with a security code. Using reverse engineering, in particular, disassemble, using a disassembler, in which direct machine code program is read and understood in its pure form, only using a mnemonic machine language, it is possible to change, modify, and even delete the security code.

As a result, a simple analysis of the state of affairs in the field of software code, it is clear that all existing protection areas are futile in terms of achieving a guaranteed result. To overcome such difficulties, the author suggests a fundamentally new technique based on a combination of virtual machine and cryptography.

Virtual machine (en virtual machine) - is a software or hardware environment, performing some code (e.g. bytecode threaded code, p-code or machine code of real processor), or the specification of such a system [5].

The main essence of the idea - the concept of a virtual machine, performing a threaded code, combined with its dynamic encryption / decryption on-the-fly (OTFE) (Fig. 2).

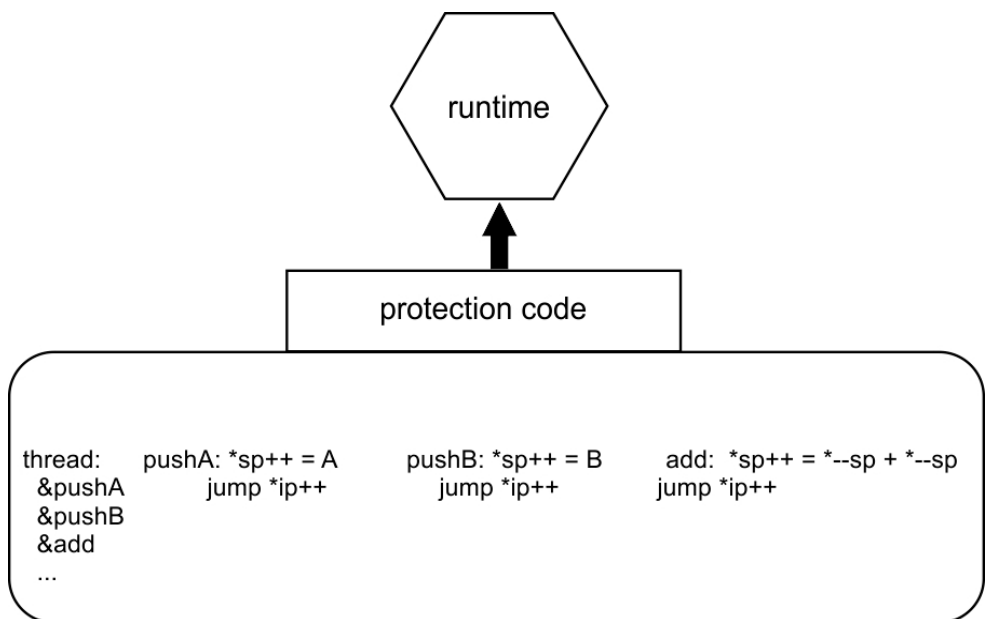

**Figure. 2. A block diagram of a virtual machine with a dynamic encryption OTFE**

Threaded code - one way of intermediate virtual machine implementation in the interpretation of programming languages (along with byte -code).

The main idea of the program by using a threaded code - an array of subroutine calls (Fig. 2). The implementation of threaded code, the method of storage of these calls may be different. This code can be processed by the interpreter (which was confirmed in the name of the address interpreter), or may be a simple sequence of machine instructions subroutine call. Some set of basic routines virtual machine using threaded code are implemented as subroutines written in the usual native [6].

Thus, program code execution speed is preserved, and simultaneously its security increases. Theoretically, performing decryption just one instruction at "time" helps to protect code with probability $p = 100\%$. In other words, the computer memory at a given time ti will be one instruction $k_i$. This will make the reversed engineering of the code impossible.

The above described technology can be implemented in different variations. As an example, consider the following sequence of steps:

1. As the encryption code key can act a certain license (key file), issued to the user (an unauthorized copying is limited).

2. The entire protected program code contains a security code fragments, dispersed by using a pseudorandom sequence generator that executes VMs , directly embedded in the main program.

As a result of these actions the original program code is "mixed" with the security code, which, in turn, is represented as a sequence of encrypted data, deciphering of which for performing is made according to one instruction , and as a result, disassembling of the security code is not possible. In this case , although the disassembling of machine code programs is still available to attacker, but does not give the desired result, because separation of the original code from the embedded security code becomes much time consuming task, not always with predictable results.

To achieve even higher degree of protection the partial implementation of the program in a virtual machine protection subsystem is possible, for example, take this code 5-10% of the total program taking 1-2 % runtime - it will not lose speed, but significantly strengthen hacking, reducing it almost to nothing (according to the particular case of the empirical law of Pareto, for 80 % of the time , the processor performs 20% of the total sold in the program commands) .

To further complicating of the analysis code one can apply the so-called "method of mixing" code. Let the command of virtual machine consists of the command code and a pointer to the next command to be processed (while the command code can be encrypted using a key as the current and / or next pointer):

```
struct command
{
CODE * code;
struct command * next_command;
}
```

Thus, the virtual machine's memory will be a linked list, which will allow to: a) mix easily the executable code , and b) effectively add and delete meaningless (so-called "junk") command. As a result, we obtain extremely simple, but very nontrivial polymorphic code, to analyze which hackers have to write a special converter. You can also implement the virtual machine memory in the form of a binary tree, pass the arguments through lists, instructions' executing organize as a deck.

Consider the most stable version of the defense - virtual machines with randomly generated set of instructions. Here any stable combination of commands can be replaced by a single compound instruction (this also minimizes the amount of executable code). If we go further, you can take a few random instructions, not even necessarily adjacent, and replace them with a new one, so continuing indefinitely. In the second generation of instructions, there would be a great choice for crushing instructions. Provided that such fragmentation will be succeeded by chance, the received instructions won't most likely coincide with the source ones, and as a result, the second generation of VM instructions will be practically unrecognizable [7].

At the same time to analyze the logic of the virtual machine is incredibly difficult. The matter is that the logical sense of random generated commands in most cases is not clear, since the number of connections between the separate commands tends to n!, Where n- number of basic instructions. For the virtual machine can handle hundreds or even thousands of commands, nor analyzing, nor disassembling of the code is possible - it all comes down to a non-standard VM algorithm of its behavior [7].

Based on the above assumptions and statements, let's make an intermediate review and present a generalized scheme of application protection (Fig. 3).

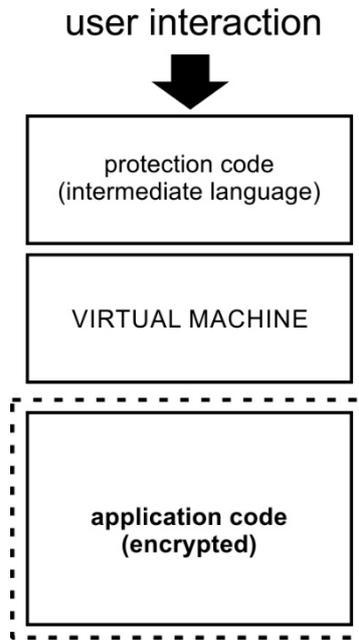

**Figure 3. Generalized scheme of application protection**

**3. Method of virtual machine in protection of software**

As you know, the perfect way to protect software does not exist, therefore, security systems developers do not seek to prevent a hacker from the possibility of security neutralization, but try to complicate this process.

Protection can solve one or a complex set of tasks such as protection from copying, illegal use, modification, etc., but despite the kind of the ultimate goal standing in front of such a product, the developers of each of them must first solve the general problem of everybody - quality protection from the study. Whatever algorithms are used for software protection, their resistance to reverse engineering determines the resistance of the entire security system as a whole.

Today in the market there are a large number of commercial protections, but many of them, including still popular ones, have been broken for a long time. Often a protection from the study brings their weak. After analyzing the intruder protection algorithms, serial keys are generated, the hardware - successfully emulated. The situation could be improved be the developing of an effective method of fixing the software protection from the study, applying which to algorithms of other protections it would be possible to raise their level of quality.

Consider the methods existing to protect software from the study:

Entanglement - artificial complication of code in order to make it more difficult readability and debugging (code mixing, implementation of false procedures, sending extra parameters in procedures, etc.)

Mutation - created every time you start up tables of operation corresponding, the operations themselves are replaced by synonyms.

Compression, encryption - the original program is packed / encrypted, and produces a reverse process as you complete.

Processors simulating - creates a virtual processor, the protected program is compiled under him, and executed on the target machine using the simulator.

There are the other methods as well as their combinations and variations, however, it's easy to see that they are all based on one simple idea: redundancy. In fact, what is obfuscated as not excessive coding program? Extra transitions, additional parameters, the extra instructions - keyword method "extra". The same applies to any of the above methods, and probably it would be natural to combine all of these methods in the same group "methods of redundant coding". What is the advantage of redundancy, in spite the fact that it increases the size of the program and reduces its speed? The matter is, that in all these protecting species is used the understanding of "human factor" - the harder the person understands the logic of a process, the more resources this process uses. For example, the functionality of one simple instruction of constant downloading to register can be "smashed" into dozens or even hundreds of instructions, and to trace the connection of all used resources (registers, memory, etc.) in this sequence is difficult for a person. The encryption method from this point of view is not unique – as in the same way as in the other methods for simple instruction (or group of ones) an abundant sequence of commands is required - in this case, these are decryption operations and plus the decrypted code operations.

However, what is automatically " confused" or complicated, it can be also automatically restored to its original state - developers of obfuscation mechanisms usually develop in parallel "unsmashing" tools, and methods of mutation and encryption imply the content of the reverse mechanism in the protected code. Alone in this group of methods is worth only method of simulation of a virtual processor, which firstly leads to a high degree of involvement and irreducible result code , and secondly (by a certain approach to implementation) , the protected code does not contain an explicit recovery methods of the original code . We consider this method in detail.

**4. Virtual processor in software protection**

The essence of the method is as follows: some functions , modules, or program entirely, are compiled under a virtual processor with an unknown potential attacker command system and architecture. The execution is provided by embedded in the resulting code simulator.

Thus, the task of reengineering of protected fragments is reduced to the study of simulator architecture , simulated processor, creating a disassembler for the latter, and, finally, the analysis of disassembling code. This task is not trivial, even for those with good knowledge and experience in working with the architecture of the target machine. Same attacker has no access to the description of a virtual processor architecture, nor to information on the organization of the simulator used. Cost of breaking increases significantly.

Why, concerning a high theoretical efficiency, this method is still not widely used? Probably for two main reasons. First, the method has features that narrow the field of its potential application - this will be discussed below. Secondly, and it's perhaps more serious reason, the complexity (and therefore the cost) of implementation of the method is very high. If we consider the fundamental possibility of information leakage about the newly created system, which immediately leads to its inefficiency and depreciation , it becomes clear why manufacturers of security software are not in a hurry to implement this method . However, it should be noted that with certain variations and limitations this method is implemented in all the latest products such as StarForce3, NeoGuard, VMProtect, etc. Apparently there will become more and more products, and existing ones will be developed , so as emerging realizations confirm the high efficiency of the method, though have yet weaknesses.

**5. Implementation of the method**

One disadvantage of this method is the high cost of implementation, but it can be considerably reduced. In the ground of the system of protection that implements this method would lie compiler of a high-level language . There is the only necessary machine-dependent phase in any compiler - code generation , the dependency in other phases , as a rule , can be eliminated .

If the compiler was originally developed as a multi-platformed, the process of migrating to a different target platform is usually simplified. For example, this can be achieved with automatic generation of code generation according to special description of the target machine. In this case, to change the platform the developers only need to modify this description. But even if there is no own compiler one can use the freely distributed open source software , for example , GCC.

And to simplify the user work with the described system of protection, it can be provided with the mechanisms ofembeddmend in popular development environments, such as MSVC. In this case, the scheme of such a complex would look like as it shown in Fig. 4.

The specific use of the compiler imposes a number of specific requirements for the virtual processor, nevertheless they can be easily implemented. There are a bit of requirements – one need only to provide access to external, according to the virtual machine memory , as well as the ability to call external functions - it is necessary for the interaction of protected and unprotected code. The rest of the architecture of a virtual processor can be completely free and the more confusing and original it'll be, the higher level of protection will be achieved.

Accordingly, the functioning in such manner secured product would occur in the scheme in Fig. 5.

The compiler itself, apart from changing cogenerating phase, need to be finalized for the acquisition the possibilities for it:
1) Distinguish access to internal and external memory relatively to the virtual processor (including function calls).
2) Create for each protected function so-called shell running on a real processor, with a call of protected function through simulation.

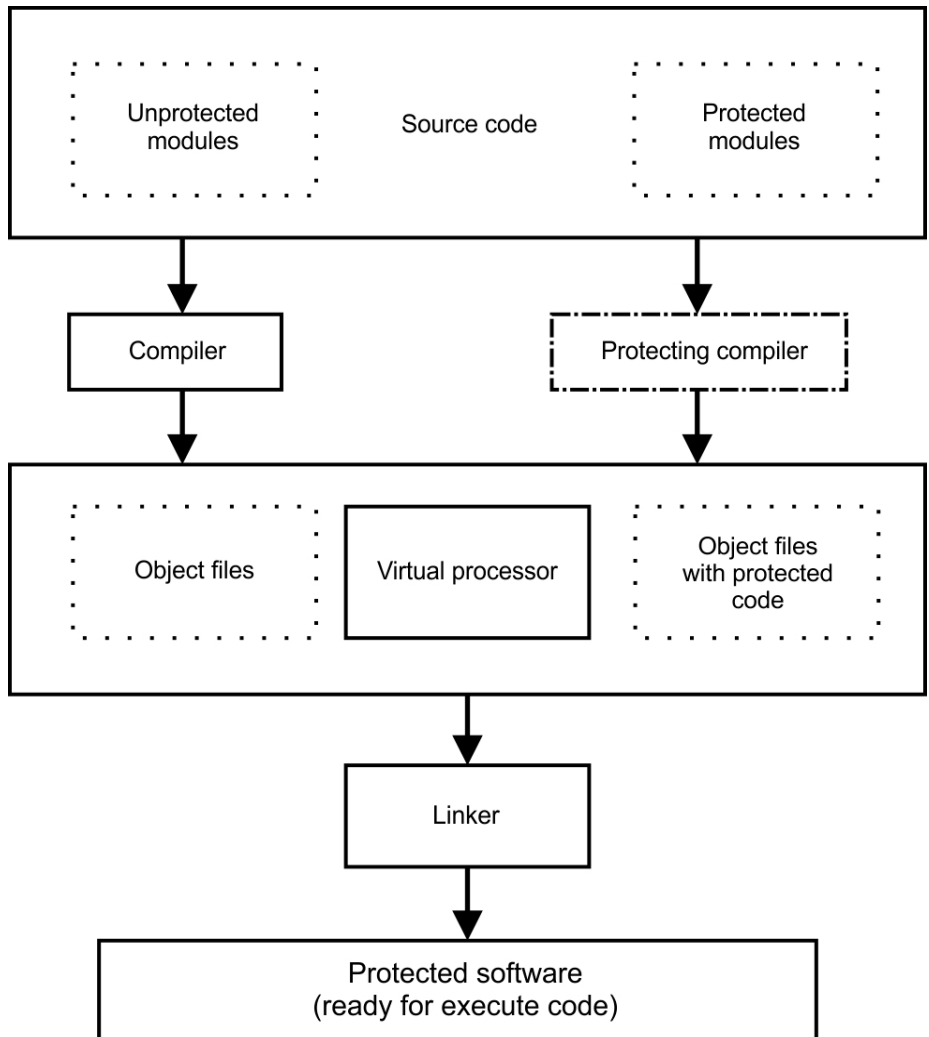
**Figure. 4. Software protection scheme**

Last opportunity should be described in detail. As reflected in the scheme in Fig. 4, functional of protected functions will be implemented through a call to the simulation, indicating which of the security functions needs interpreting. However, before that, you need to perform special code that prepares the protected function parameters - "move" them with real registers and virtual memory, the method according to the architecture of the virtual processor. All these will deal with special functions generated by our compiler – "shell". Shell, in turn, will use the special features of the simulator to access virtual registers and memory. Characteristically, our compiler will generate a shell in a high level language, which, in turn, will compile the standard compiler, used by the user to build the unprotected part of your project.

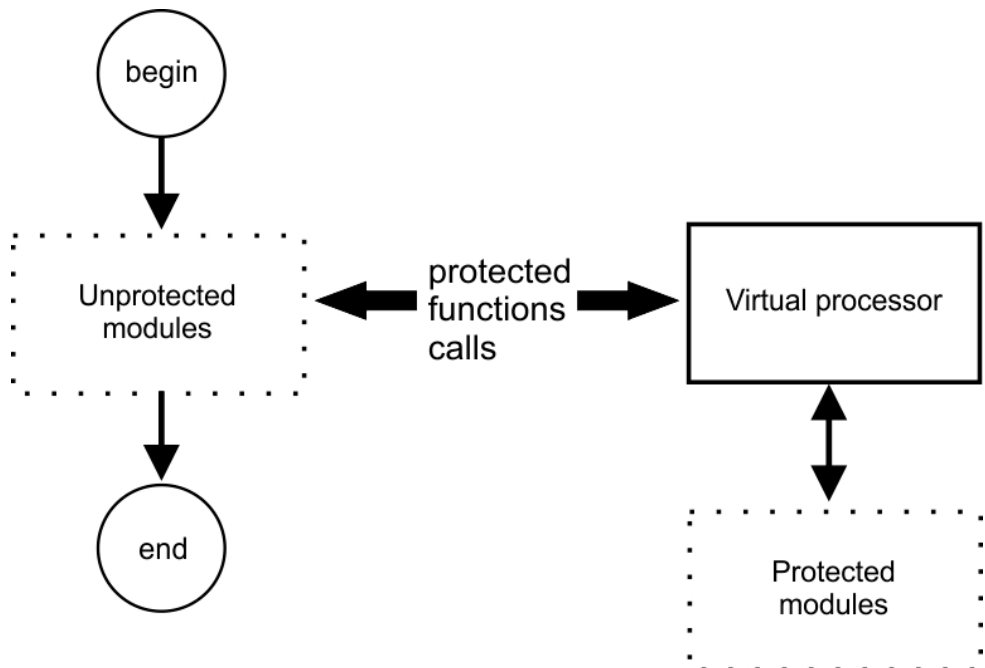
**Figure. 5. Scheme of protected software**

Of course, the specific implementation method of a virtual processor may be somewhat different from that described so as the scheme of protected product work. Nevertheless, the described variant is quite viable, and, furthermore, is relatively simple.

The heart of a protected product - a simulator. It will be included in any assembly of the protected product. But, we will not consider its implementation, as special requirements to it are not presented - it should only simulate the architecture of our virtual processor operations, including access to external memory. However, it should be noted that according to the nature of its application, it is necessary to automate maximum the adjusting process to the new virtual simulator architectures.

Disadvantages of the method are consequence of its advantages:
a) The speed of the transferred code to a virtual environment at times (approximately 10-50, depending on the architecture of a virtual processor and simulator ) is lower than the original code.
b) The volume of protected software will usually be somewhat higher than the unprotected one.

However, the last drawback is irrelevant, since the size will increase slightly, and in some cases may even decline. The first drawback is principle and imposes some obvious limitations to the use of the method.

## 6. Conclusion

A fundamentally new approach to building systems of protection of executable program code is developed. Method of software protection based on virtual machine is proposed. A high resistance to cracking the security subsystem is proved.

The considered method of software protection is very effective, given that the cost of its development can be greatly reduced. However, the features of the method can not be recommended for the protection of its programs completely. Thus, the method can not be applied for protection functions that are critical to execution time, which may slow down and considerably reduce the efficiency of use of the program by the user. Nevertheless, careful use of this method allows to achieve a very high level of protection from the study. In this connection, the main area of application is seen increasing resistance to the study of individual algorithms of the other security systems software. In addition, the method can be applied to protect the not-intensive algorithms, as well as to conceal the content of the protected program, some special data, such as information about authorship.

Potential of virtual machine technology in the field of information security is very high. To implement complex hardware protection is not always advisable in view of the cost of the time, money and complexity of implementation. Software implementation is extremely flexible and simple, limited only by means of mathematics, programming languages and, of course, professional developers.

**References**

1. Технология криптозащиты кода. / Дмитрий Козлов // http://hardline.ru
2. Список популярных дизассемблеров. // http://wasm.ru/toollist.php?list=13.
3. UPX: the Ultimate Packer for eXecutables // http://upx.sourcefourge.net.
4. http://www.aspack.com/asprotect.aspx
5. http://ru.wikipedia.org/wiki/Виртуальная_машина
6. http://ru.wikipedia.org/wiki/Шитый_код
7. Петров А.С. Технология защиты программного кода посредством применения виртуальной машины. / Петров А.С., Петров А.А. // Вестник ВНУ № 9(103), часть 1. с. 117-122.
8. Метод виртуального процессора в защите программного обеспечения // http://citforum.ru/security/software/virt_proc